\documentclass[aps,prl,twocolumn,amsmath,amssymb,amsfonts]{revtex4-1}
\usepackage[varg]{txfonts}
\usepackage{times}
\usepackage{amsmath}
\usepackage{color}
\usepackage{graphicx}
\usepackage[percent]{overpic}
\usepackage{wasysym}
\usepackage{caption}
\usepackage{subcaption}

\newcommand{\bra}[1]{\left \langle #1\right|}

\newcommand{\ket}[1]{\left| #1  \right \rangle}

\newcommand{\avg}[1]{\langle #1 \rangle}

\newcommand{\h}[1]{{#1}^{\dagger}} 
\newcommand{\cc}[1]{{#1}^{*}}
\newcommand{\cb}[1]{\bar{#1}}

\definecolor{acolor1}{RGB}{32,90,0}
\definecolor{acolor2}{RGB}{228,0,45}
\definecolor{acolor3}{RGB}{69,49,215}

\begin{document}

\title{Symmetry breaking via Kondo hybridization: Chiral nematic metal in Pr$_{\bf 2}$Ir$_{\bf 2}$O$_{\bf 7}$}
\author{Jeffrey G. Rau}
\affiliation{Department of Physics, University of Toronto, Toronto, Ontario M5S 1A7, Canada}
\author{Hae-Young Kee}
\email[Electronic Address: ]{hykee@physics.utoronto.ca}
\affiliation{Department of Physics, University of Toronto, Toronto, Ontario M5S 1A7, Canada}
\affiliation{Canadian Institute for Advanced Research/Quantum Materials Program, Toronto, Ontario MSG 1Z8, Canada}

\date{\today}
 \begin{abstract}
   In frustrated magnets when
   magnetic ordering is
   suppressed down to low temperature, 
   the formation of a quantum
   spin liquid becomes a possibility.
   How such a spin
   liquid manifests in the presence of conduction electrons is a question
   with potentially rich physical consequences, particularly
   when both the localized spins
   and conduction electrons reside on frustrated lattices.
   We propose a novel mechanism for symmetry breaking
   in systems where conduction electrons hybridize with
   a quantum spin liquid through Kondo couplings.   
   We apply this to the pyrochlore iridate
   Pr$_2$Ir$_2$O$_7$, which exhibits an anomalous Hall effect without clear indications of magnetic order.
   We show that Kondo hybridization between the localized Pr pseudo-spins
   and Ir conduction electrons breaks some of the spatial symmetries, in addition to time-reversal
   regardless of the form of the coupling. These broken symmetries result in  
   an anomalous Hall conductivity and induce small
   magnetic, quadrupolar and charge orderings. Further experimental signatures are proposed.
\end{abstract}

\maketitle

\emph{Introduction:} The study of interactions between itinerant electrons and localized
degrees of freedom has lead to an understanding of a wealth of novel physical 
phenomena. These range from isolated moments, as in
the Kondo effect\cite{kondo1964resistance,wilson1975renormalization} through into the
realm of dense lattices of moments as in heavy fermion materials\cite{stewart1984heavy,doniach1977kondo,hewson1997kondo}
and the anomalous Hall effect (AHE)\cite{nagaosa2010anomalous}.
While still largely unexplored, the interplay 
between itinerant degrees of freedom and frustrated local
moments promises to unveil new and unique phases of matter\cite{si2006global}.
One particularly interesting scenario arises when the local moments are highly frustrated, 
realizing a quantum spin liquid. How such a spin liquid competes with Kondo hybridization
when conduction electrons are present has yet to be fully addressed 
\cite{senthil2003fractionalized,senthil2004weak,ghaemi2007higher,coleman2010frustration}.

\begin{figure}[tp]
\begin{subfigure}[b]{0.45\columnwidth} 
  \begin{overpic}[width=\textwidth]
    {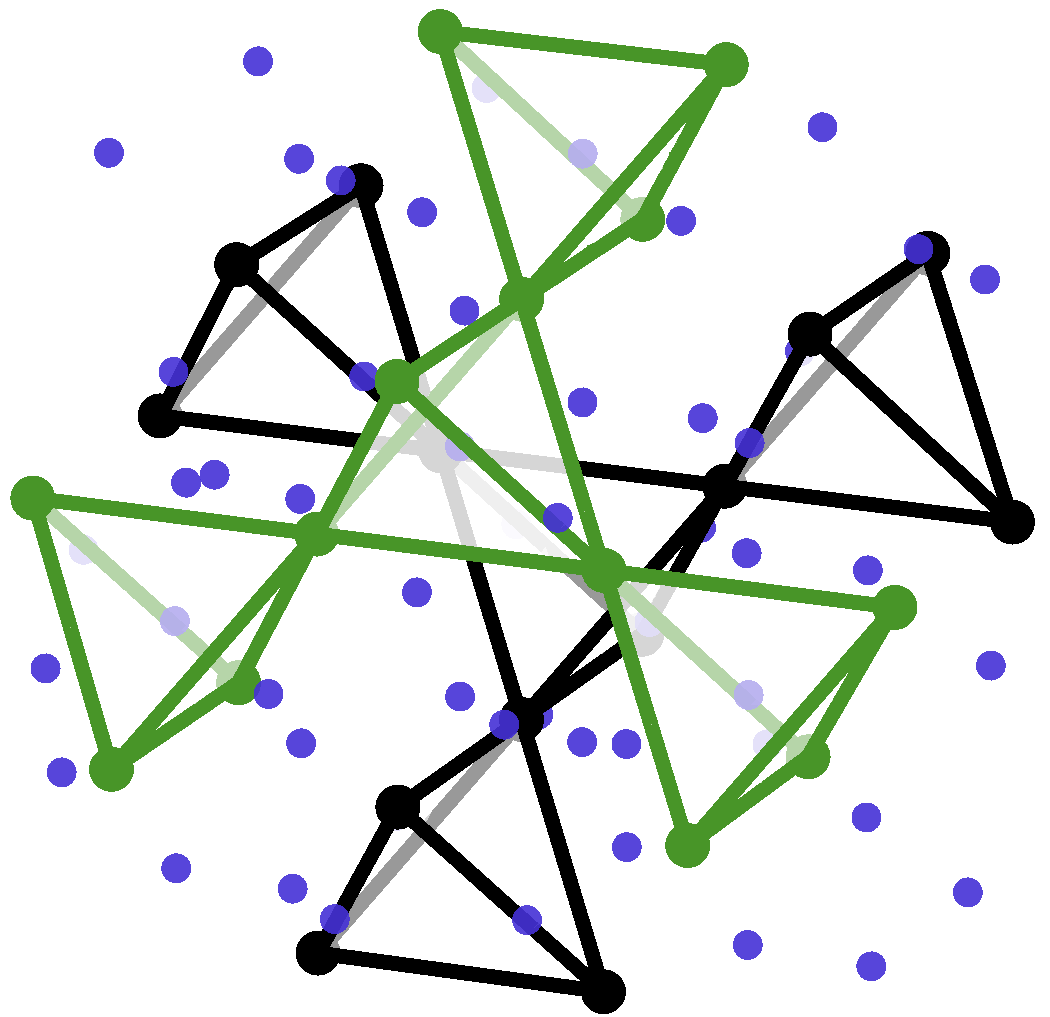}
    \put(20,78){\small {\fontfamily{ptm}\selectfont Ir}}
    \put(16,20){\textcolor{acolor1}{\small {\fontfamily{ptm}\selectfont Pr}}}
    \put(80,88){\textcolor{acolor3}{\small {\fontfamily{ptm}\selectfont O}}}
  \end{overpic}
  \caption{  \label{structure}}
\end{subfigure}
\begin{subfigure}[b]{0.52\columnwidth}
  \begin{overpic}[width=\textwidth]
    {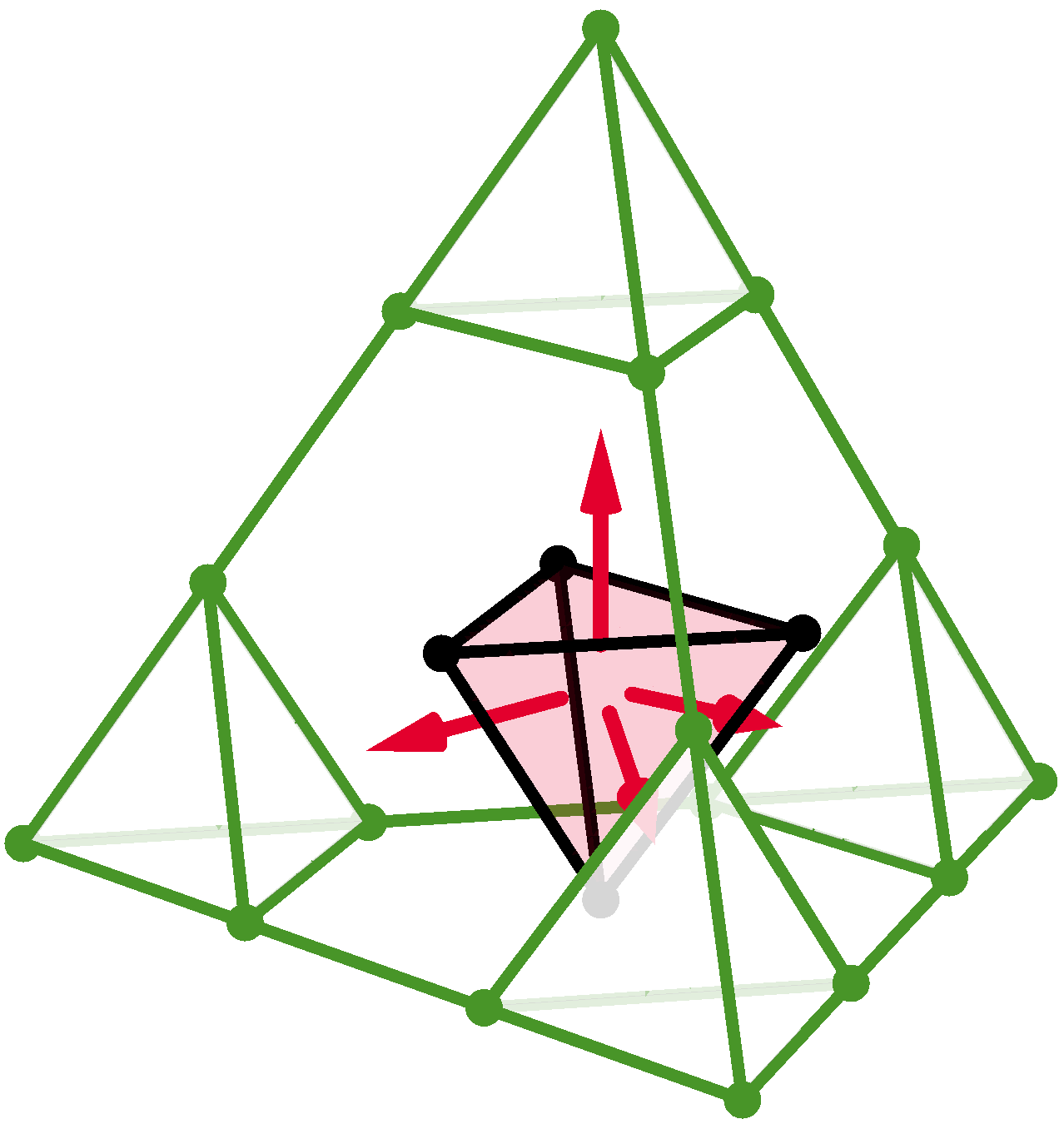}
    \put(70,48){\small {\fontfamily{ptm}\selectfont Ir}}
    \put(20,10){\textcolor{acolor1}{\small {\fontfamily{ptm}\selectfont Pr}}}
    \put(45,58){\textcolor{acolor2}{\small {\fontfamily{ptm}\selectfont $\frac{\pi}{2}$}}}
  \end{overpic}
  \caption{ \label{fluxes}}
\end{subfigure}
\caption{
\label{pattern1} \small (a) Crystal structure of Pr$_2$Ir$_2$O$_7$. (b) Illustration of flux pattern induced from the local basis rotation, with spin-dependent parts ignored for
clarity.
}
\end{figure}
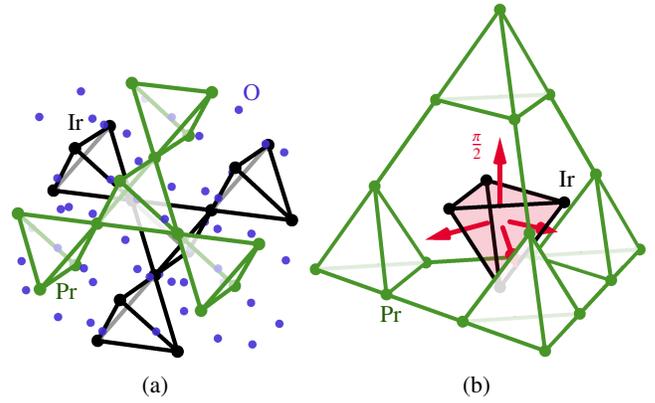

In this letter, we study systems where conduction electrons interact with
a quantum spin liquid, introducing a novel mechanism for breaking
spatial symmetries. When the conduction electrons hybridize
with spinons the emergent gauge
structure of the spin liquid is exposed. We propose that a spin liquid with non-trivial 
gauge structure, i.e. fluxes penetrating the lattice,
is incompatible with trivial gauge structure in the conduction
states as well as some of the spatial symmetries.
We apply this to a model of conduction electrons
and local moments on the pyrochlore lattice, where the effective fluxes are 
provided by choosing local quantization axes for the conduction electrons. While this
is simply a basis choice when the electrons are isolated,
when coupled with a fully symmetric ${\rm U}(1)$ spin liquid on the local moments
any uniform hybridization forces the emergent magnetic flux
through the plaquettes between the local 
moments and the conduction electrons (as shown in Fig. \ref{fluxes})
breaking some of the spatial symmetries.

A puzzling example of a material with frustrated
local moments and conduction electrons arises in the pyrochlore iridiate Pr$_2$Ir$_2$O$_7$,
where the Praeseodymium (Pr)
and Iridium (Ir) atoms form of a pair of interpenetrating pyrochlore lattices with space group $Fd\cb{3}m$,
as shown in Fig. \ref{structure}.
The lack of
indications of magnetic ordering\cite{maclaughlin2009weak} well below the Curie-Weiss temperature\cite{nakatsuji2006metallic,machida2007unconventional} suggests that
the Pr sublattice is frustrated, either intrinsically or 
due the presence of the Ir conduction electrons\cite{zhou2008dynamic}. This is corroborated by features in the
field dependent magnetization at low
temperatures suggesting an anti-ferromagnetic interaction
and possibly spin-ice physics, in contrast
to the sign of the Curie-Weiss temperature\cite{machida2009time}. In addition to these
magnetic features, the compound is metallic\cite{nakatsuji2006metallic},
originating in the Ir sublattice, and shows a finite AHE at intermediate temperatures between
$\sim 0.3K$ and $\sim 1.5K$\cite{machida2009time}. The presence of an AHE along $[111]$ indicates
a breaking of time-reversal symmetry as well the rotational symmetry of the lattice.

Generically, since magnetization and the anomalous Hall vector $\vec{\sigma}_A = \sigma_{yz}\hat{x}+\sigma_{zx}\hat{y}+\sigma_{xy}\hat{z}$ transform
in an identical fashion one expects the two orderings to appear together, as is
found in ferromagnets\cite{nagaosa2010anomalous}. The
mystery in Pr$_2$Ir$_2$O$_7$ is that the intermediate phase shows 
no evidence for net magnetization to a resolution of $\sim 10^{-3} \mu_B / {\rm Pr}$\cite{machida2009time}. A number of other
unconventional features, such as the lack of a clear phase transition into this intermediate phase
as well as unusual behaviour of the Hall conductivity in large fields\cite{nakatsuji2006metallic,machida2007unconventional,machida2009time},
further enrich the problem. 
These unexplained properties have attracted considerable theoretical attention, with
proposals exploring the full range of scenarios
from the interplay between spin-ice physics and the conduction electrons\cite{machida2009time,udagawa2012anomalous}
to detailed considerations of the Ir physics\cite{moon2012non,flint2013chiral} and Pr-Ir couplings\cite{chen2012magnetic,lee2013rkky}. 
While a consensus has yet to emerge, it is clear that both the Pr and Ir degrees of freedom must be taken into
account to explain the fascinating phenomena seen in
experiments. We employ idea to understand Pr$_2$Ir$_2$O$_7$, finding that
a uniform ${\rm U}(1)$ spin liquid is favoured, leading to breaking of the appropriate
symmetries to allow an AHE when hybridization is included. The orbital nature
of this symmetry breaking provides a simple explanation for
both the AHE as well as the smallness of the induced magnetic 
and quadrupolar moments.

\emph{Conduction electrons:} We first construct a minimal model for Pr$_2$Ir$_2$O$_7$, beginning with the
Ir atoms. Assuming an ionic configuration of Ir$^{4+}$ one has
five $d$ electrons per Ir. These Ir$^{4+}$ ions form
a pyrochlore lattice, face centered cubic with a tetrahedral basis, each 
surrounded by oxygens.  Due to the strong octahedral crystal
fields and spin-orbit coupling, one can consider
only a single half-filled $j_{\rm eff}=1/2$ band\cite{kim2008novel}.
In the global cubic axes
a symmetry operation $S$ rotates the spin and orbital degrees of freedom
according to some representation $R_S$. 
How these symmetry operations
act with the local axes can be seen most clearly if we adopt quantization axes 
for the $j_{\rm eff}=1/2$ states
that are compatible with the exact $D_{3d}$ site symmetry of the Ir$^{4+}$ ions. 
These axes are defined so the $\hat{z}$ axis points along the
local $[111]$ direction and the $\hat{y}$ axis is oriented along
 one of the $C'_2$ axes perpendicular to the local $[111]$, with
frames on different basis sites related by $C_2$ rotations.
 If we consider rotations of the $d$ levels $U_r$ at each site $r$
that take the global cubic axes to the local frames then the operation
$S$ acts in the local frame as $U_{S(r)} R_S \h{U}_r$. The set of quantization axes
for the pyrochlore lattice
has the advantage of acting only in the local frames, with the rotations
of the local spin being the same across all the basis sites of the 
lattice up to a sign. Explicitly,  one finds
\begin{equation}
  \h{P} U_{S(r)} R_S \h{U}_r P = z_{S,r} L_S,
\end{equation}
where the operator $P$ projects
into the $j_{\rm eff}=1/2$ subspace of the $d$ levels. 
The $z_{S,r}$ is a sign that only depends on the basis site of
the pyrochlore lattice and can be found in Ref. \cite{burnell2009monopole}
as the gauge transformations for the monopole flux state.
The $L_S$ are spin rotations in the $\Gamma_{j=1/2}=\Gamma_{4g}$ representation
of the site symmetry group $D_{3d}$ and be obtained from the generators
\begin{eqnarray}
  &&L_{C_3} = e^{-i \pi \sigma^z/3}, \ \ \ \ L_{C_2} =  L_{I} = 1, \\
  &&L_{C'_2} = L_{C_4} = i\sigma^y
\end{eqnarray}
where $C_3$ and $C_2$ are independent of axis and 
the $C'_2$ and $C_4$ are for the $[110]$ and $[100]$ axis respectively.

Here we will work only with the nearest
neighbour hoppings, where aside from spin
the hopping matrices depend only on the four basis sites.
Extension to further neighbour hoppings is straightforward.x
Using symmetry operations in the local axes,
$\h{c}_r \rightarrow z_{S,r} L_S \h{c}_{S(r)}$, one can show that there are
only two allowed terms in the model
\begin{equation}
H_{\rm Ir} = \sum_{\avg{rr'}} i \h{c}_r\left[t_1 \sigma^z \gamma^z_{rr'} + t_{2}(\sigma^+ \gamma^+_{rr'} + \sigma^- \cb{\gamma}^+_{rr'})\right]c_{r'}.
\end{equation}
The $\gamma^z_{rr'}$ and $\gamma^{+}_{rr'}$ depend only the basis sites and can
be written
\begin{equation*}
   \gamma^+ = 
  \left(
  \begin{tabular}{cccc} 
    $0$ & $+1$ & $+\cb{\omega}$ & $+\omega$ \\
    $-1$ & $0$ & $+\omega$ & $-\cb{\omega}$ \\
    $-\cb{\omega}$ & $-\omega$ & $0$ & $+1$ \\
    $-\omega$ & $+\cb{\omega}$ & $-1$ & $0$ 
  \end{tabular}
  \right), \ \ \
   \gamma^z = 
  \left(
  \begin{tabular}{cccc}
    $0$ & $+1$ & $+1$ & $+1$ \\
    $-1$ & $0$ & $+1$ & $-1$ \\
    $-1$ & $-1$ & $0$ & $+1$ \\
    $-1$ & $+1$ & $-1$ & $0$ 
  \end{tabular}
  \right),
\end{equation*}
where $\omega = e^{2\pi i/3}$.  Earlier studies have used formal global axes for
the $j_{\rm eff}=1/2$ bands\cite{kurita2011topological,krempa2013pyrochlore}, which
can be obtained from the model derived above by inverting the local spin rotation
$U_r$ at each site.

\emph{Non-Kramers doublets and pseudo-spins:} Having established a model for the Ir$^{4+}$ ions, we now consider the 
Pr$^{3+}$ ions. Since these states are highly localized, being in a $4f^2$ configuration,  
we use Hund's rules to arrive at the ground state multiplet
${}^3H_4$, with inelastic neutron
scattering studies of Pr$_2$Ir$_2$O$_7$ identifying a ground state
doublet of $E_g$ character.
The lowest lying excited state is a singlet $\sim 162K$\cite{machida2005crystalline} above the doublet,
two orders of magnitude larger
than the onset of the ordering, so we restrict to only the ground state
doublet. This doublet has the form
\begin{equation}
 \ket{E_g,\pm} = a_4 \ket{\pm 4} \pm a_1 \ket{ \pm 1} -a_2 \ket{\mp 2},
\end{equation}
where $a_4$, $a_2$ and $a_1$ are real numbers depending on the details
of the crystal field\cite{onoda2010quantum}.  
Within the space of doublets, super-exchange interactions are mediated
through the surrounding oxygen atoms. This can be computed via
a strong coupling expansion, including the effects of hopping
between the Pr $4f$ states and the O $2p$ states.
When projected into the subspace of doublets,
the exchange Hamiltonian is most conveniently written using
pseudo-spin operators
\begin{equation}
  \tau^\mu_r = \sum_{\alpha \beta} \ket{E_g,\alpha}_r\bra{E_g,\beta}_r \sigma^\mu_{\alpha \beta},
\end{equation}
where $\alpha,\beta=\pm$, $\mu = x,y,z$ and $\ket{E_g,\pm}_r$ are the doublet states at site $r$. 
The $\tau^z_r$ operator is magnetic, proportional to the
magnetic dipole moment, while the transverse $\tau^x_r$ and $\tau^y_r$ parts are non-magnetic,
carrying quadrupolar moments.
All three exchanges allowed by symmetry are generated\cite{footnote1},
giving the model in the local axes\cite{onoda2010quantum,onoda2011quantum,lee2012generic},
\begin{eqnarray}
H_{\Pr} &= &\sum_{\avg{rr'}}\left[
J_z \tau^z_r \tau^z_{r'} + \frac{J_\perp}{2}\left(\tau^+_r\tau^-_{r'} +\tau^-_{r} \tau^+_{r'}\right)\right] \\
& +& J_{\pm \pm}\sum_{\avg{rr'}}\left(\gamma_{rr'} \tau^+_r\tau^+_{r'} +\cb{\gamma}_{rr'} \tau^-_r \tau^-_{r'} \right),
\end{eqnarray}
where  $\tau^{\pm}_r = \tau^x_r \pm i\tau^y_r$ and 
the sums run over nearest neighbour bonds. 
Pseudo-spin rotational symmetry is not present when
$J_z \neq J_{\perp}$ or in the presence of $J_{\pm\pm}$.
The form of the $J_{\pm\pm}$ terms is a consequence of the intertwining of
pseudo-spin and spatial symmetries, with the phases
$\gamma_{rr'}$ defined as $\gamma_{rr'} = \cb{\gamma}^+_{rr'} \gamma^z_{rr'}$.
\begin{figure}[tp]
\begin{subfigure}[b]{0.43\columnwidth} 
  \begin{overpic}[width=\textwidth]
    {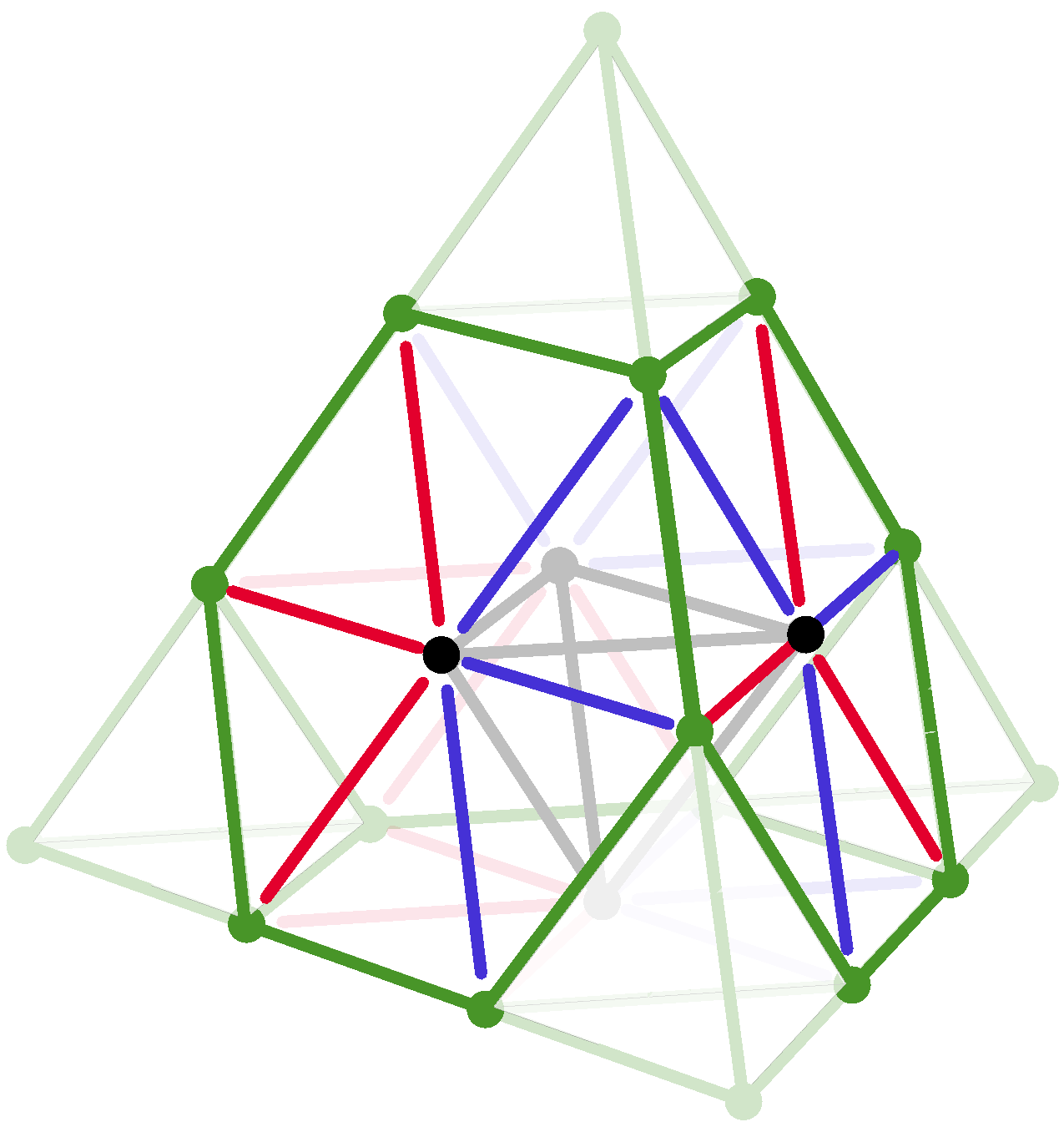}
    \put(31,46){\small {\fontfamily{ptm}\selectfont Ir}}
    \put(21,8){\textcolor{acolor1}{\small {\fontfamily{ptm}\selectfont Pr}}}
  \end{overpic}
  \caption*{}
\end{subfigure}
\begin{subfigure}[b]{0.52\columnwidth} 
\begin{subfigure}[t]{\columnwidth} 
\begin{subfigure}[b]{0.47\textwidth} 
  \begin{overpic}[width=\textwidth]
    {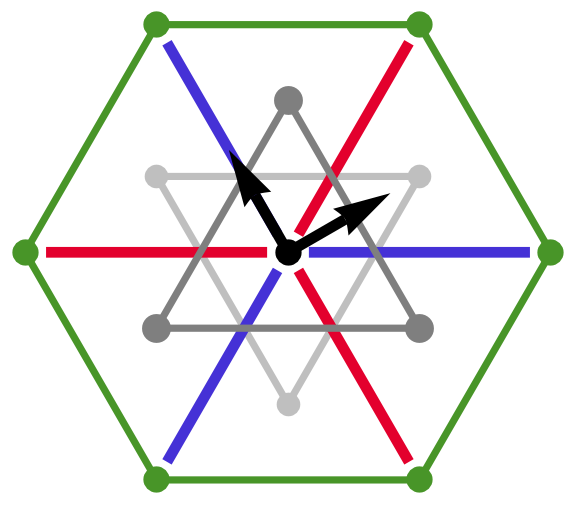}
    \put(65,47){\tiny $\hat{x}$}
    \put(33,53){\tiny $\hat{y}$}
  \end{overpic}
  \caption*{\label{hexagon-1}$[111]$}
\end{subfigure} 
\begin{subfigure}[b]{0.47\textwidth} 
  \includegraphics[width=\textwidth]{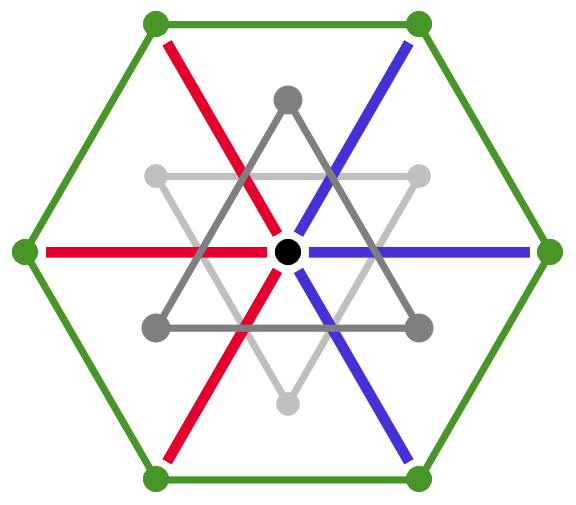}
  \caption*{\label{hexagon-2}$[1\cb{1}\cb{1}]$}
\end{subfigure} 
\end{subfigure} 
\begin{subfigure}[t]{\columnwidth} 
\begin{subfigure}[b]{0.47\textwidth} 
  \includegraphics[width=\textwidth]{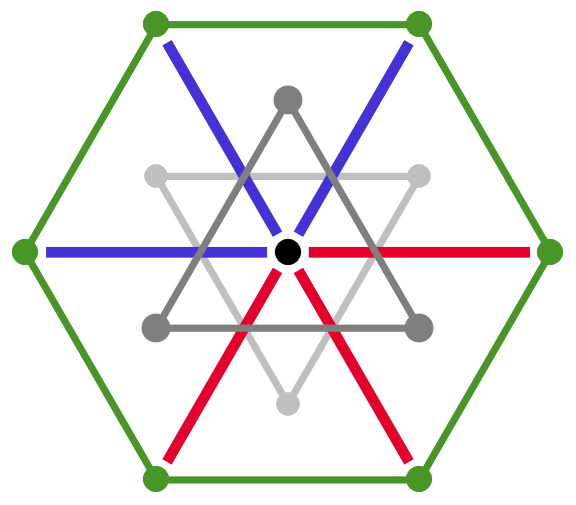}
  \caption*{\label{hexagon-3}$[\cb{1}1\cb{1}]$}
\end{subfigure} 
\begin{subfigure}[b]{0.47\textwidth} 
  \includegraphics[width=\textwidth]{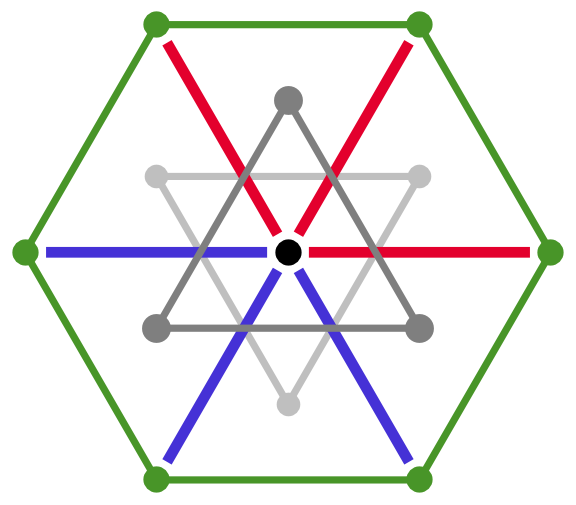}
  \caption*{\label{hexagon-4}$[\cb{1}\cb{1}1]$}
\end{subfigure} 
\end{subfigure} 
\end{subfigure}
\caption{\label{pattern2}\small Hybridization form factor
 $(-1)_{r,r'}\gamma^z_{rr'}$ for $\Gamma_{5u}$ and $\Gamma_{6u}$
 intermediate state, where $+1$ is
shown in blue and $-1$ in red. Explicit form factors
for the hexagon in each $[111]$ plane are shown alongside,
oriented so the local $\hat{x}$ axis at the central
site is $60^\circ$ from the vertical.}
\end{figure}

\emph{Hybridization:} We now consider interactions between the Pr and Ir sublattices, focusing on
those mediated by hoppings between the sublattices, through physical or virtual processes.
Charge transfer between the Pr and Ir necessarily involves 
intermediate states such as $4f^1$ or $4f^3$. For
definiteness, we will assume that the $4f^1$ states
are lower in energy than the $4f^3$, and thus dominate, though our
results do not depend fundamentally on this choice. In the $D_{3d}$
crystal field this splits into a combination of $\Gamma_{4u}$,
$\Gamma_{5u}$ and $\Gamma_{6u}$ representations\cite{bradley1972mathematical}. An example is the pair $\Gamma_{5u}+\Gamma_{6u}$,
degenerate due to Kramers theorem, given by the $m = \pm 3/2$
states in $j=5/2$ manifold of the $4f^1$ configuration. Hybridization between
the the $5d$ $j_{\rm eff}=1/2$ states of the Ir and the localized states
on the Pr can occur via several mechanisms, such as
oxygen mediated hoppings, but an effective description written as
direct hopping is possible once the intermediate states have been
integrated out. Considering only intermediate states
$\Gamma_{5u}$ and $\Gamma_{6u}$, the allowed hoppings are
\begin{eqnarray*}
H_{\rm hyb} &=& V_z \sum_{rr'} \gamma^z_{rr'} e^{i\pi
\alpha/4}(-1)_{r,r'} \h{c}_{r\alpha} \ket{\Gamma_{5u}}_{r'}\bra{E_g,\cb{\alpha}}_{r'}\\
&+& V_{\pm} \sum_{rr'} \gamma^{\alpha}_{rr'} e^{i\pi\alpha/4} (-1)_{r,r'}\h{c}_{r\alpha} \ket{\Gamma_{5u}}_{r'}\bra{E_g,\cb{\alpha}}_{r'}\\
&+&  {\rm time\ reversed}+ {\rm h.c.}
\end{eqnarray*}
where $\cb{\alpha}=-\alpha$, $r$ is an Ir site, $r'$ is a Pr site and $\gamma^{-}_{rr'} = \cb{\gamma}^{+}_{rr'}$. If
one splits the Ir-Pr bonds into two sets, related
by inversion, then $(-1)_{r,r'}$ is $+1$ on
the first set and $-1$ on the second. This pattern is shown
for the Pr centered hexagons in Fig. \ref{pattern2}.
 To derive this form,
one must keep in mind that the $\Gamma_{5u}$
and $\Gamma_{6u}$ states are Kramers and defined in
the local axes, and so carry the same signs $z_{S,r}$ as the $j_{\rm eff}=1/2$ states
in their symmetry operations. For simplicity
we set $V_{\pm}=0$ for the remainder of this work, as
it does not affect the results qualitatively. 

We consider a fermionic slave-particle approach, as this
allows for a natural treatment of hybridization between the Pr and Ir.
The transition operators $\ket{\Gamma_{5u}}_r\bra{E_g,\alpha}_r$
and $\ket{\Gamma_{6u}}_r\bra{E_g,\alpha}_r$ are written using a pseudo-spinon $\eta_{r\alpha}$ and auxiliary bosons $\Phi_5$ and $\Phi_6$
\begin{eqnarray}
  \ket{\Gamma_{5u}}_r \bra{E_g,\alpha}_r &=& \h{\Phi}_{5,r} \eta_{r\alpha}, \\
  \ket{\Gamma_{6u}}_r \bra{E_g,\alpha}_r &=& \h{\Phi}_{6,r} \eta_{r\alpha} 
\end{eqnarray}    
  These slave-particles are constrained
to satisfy $\h{\eta}_r \eta_r +\h{\Phi}_{5,r} \Phi_{5,r}+\h{\Phi}_{6,r} \Phi_{6,r}=1$.
Since these pseudo-spinons are of non-Kramers character,
the symmetry operations in this local basis do not carry the signs
$z_{S,r}$ and transform simply as $\h{\eta}_r \rightarrow M_S \h{\eta}_r$ where $M_S$ is the pseudo-spin rotation
corresponding to the symmetry operation $S$. The $\Phi_5$ and $\Phi_6$
bosons transform as the associated one-dimensional representations,
but being Kramers states in the local quantization
axes they also carry the phase factors $z_{S,r}$ and transform as 
$\Phi_{5,r} \rightarrow z_{S,r} e^{i\phi_{5,S}} \Phi_{5,S(r)}$ and $\Phi_{6,r} \rightarrow z_{S,r} e^{i\phi_{6,S}} \Phi_{6,S(r)}$ under the symmetry operation $S$.

When splitting $\Delta$ between the $E_g$ and the 
excited states is large then we expect
condensation of the bosons $\Phi_{5,r}$ 
and $\Phi_{6,r}$ at order $\Delta^{-1}$. In this limit the constraint
can be simplified to $\h{\eta}_r \eta_r \sim 1$.
Condensing only in the $\Phi_5$ channel one has an
effective hopping between electron $\h{c}_r$ and spinon $\eta_{r'}$ 
\begin{equation}
  \label{hyb1}
  H_{\rm hyb} \sim V \sum_{rr'} \gamma^z_{rr'} e^{i\pi
\alpha/4}(-1)_{r,r'} \h{c}_{r\alpha}  \eta_{r'\cb{\alpha}} +{\rm h.c.}
\end{equation}
where we have absorbed $\cc{\Phi}_5$ into $V_z$ defining $V \equiv V_z \cc{\Phi}_5$.
Having either the $\Phi_5$ or $\Phi_6$ channels to condense
breaks time-reversal and time-reversal squared, an example of
hastatic order\cite{chandra2013hastatic}. 
However the one-dimensional nature of $\Gamma_{6u}$ and $\Gamma_{6u}$ allows
the combination $H_{\rm Ir} + H_{\rm hyb}$ to break none of the spatial symmetries 
of the problem, a key difference from the case considered in Ref. \onlinecite{chandra2013hastatic}.
\begin{figure}[tp]
\begin{subfigure}[b]{0.49\columnwidth} 
  \includegraphics[width=\textwidth]{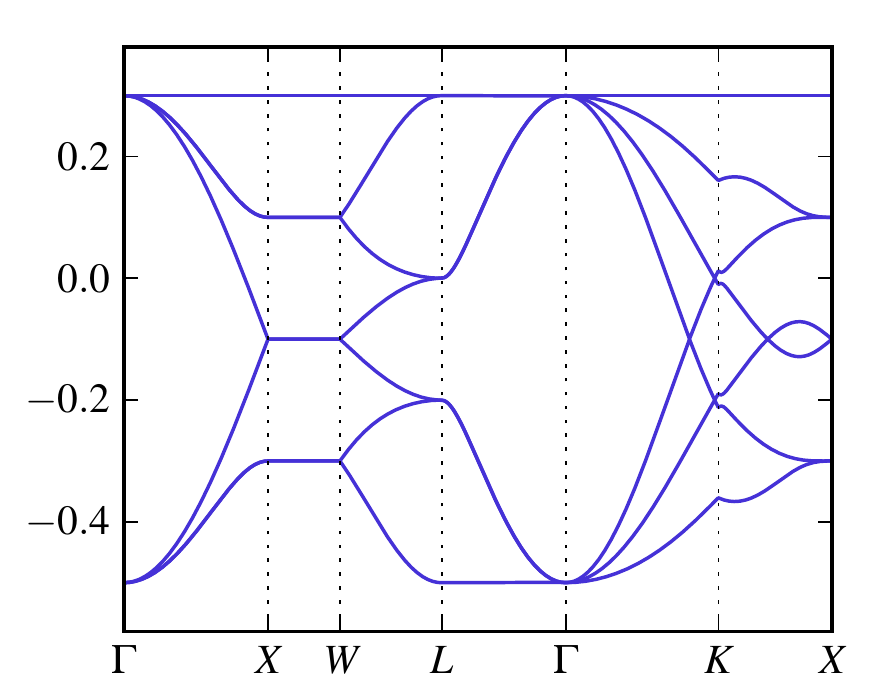}
  \caption{\label{uniform-bands-1}$E = 2\chi$}
\end{subfigure}
\begin{subfigure}[b]{0.45\columnwidth} 
  \includegraphics[width=\textwidth]{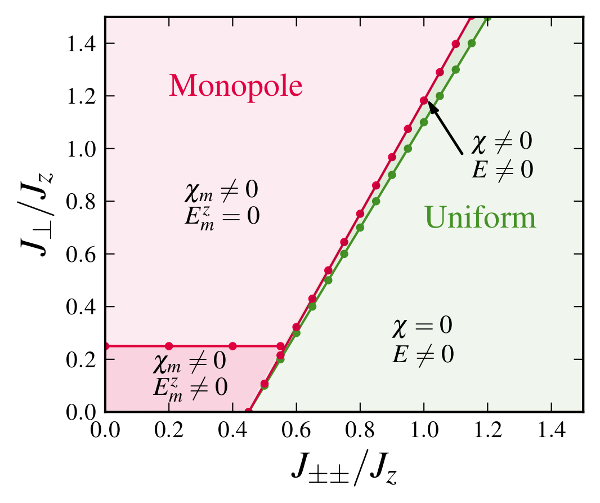}
  \caption{\label{phase-diagram}Phase diagram}
\end{subfigure}
\caption{\small
(a) The band structure of the uniform ansatz for $E=2\chi = 0.1$.
(b) The phase diagram considering triplet extensions to both the monopole and uniform ansatzes.}
\end{figure}

\emph{Non-Kramers spin liquids:} In terms of the slave-particles the pseudo-spin
operator is given by $\tau^\mu_r = \frac{1}{2}\h{\eta}_r \sigma^\mu \eta_r$. To render 
the problem tractable, we consider an approximate ground state generated from
a Hamiltonian quadratic in the fermions.
Variational Monte Carlo calculations\cite{kim2008chiral,burnell2009monopole}
on the Heisenberg model motivate us to consider two classes
of $U(1)$ spin liquid ansatzes, the uniform and monopole states which 
are competitive in this limit.
The monopole ansatz is a chiral spin liquid, breaking time-reversal and inversion but preserving the product,
and can be characterized by hoppings carrying a flux of $\pi/2$ exiting the faces of each tetrahedron.
The uniform state has equal hoppings on all bonds, carrying zero flux through all plaquettes.
Since the presence of $J_{\pm\pm}$ or $J_{z} \neq J_{\perp}$ breaks ${{\rm SU}}(2)$ pseudo-spin rotational symmetry,
these ansatzes must be extended using 
their respective projective symmetry group to include pseudo-spin-dependent $E^{\alpha}_{rr'}$ hoppings
in addition to the pseudo-spin-independent $\chi_{rr'}$ hoppings allowed at the ${{\rm SU}}(2)$
symmetric point. 
Each spin liquid ansatz is characterized by a quadratic Hamiltonian
\begin{equation*}
\label{mf-hamiltonian}
H(\chi,E) = \sum_{\avg{rr'}}\left( \chi_{rr'} \h{\eta}_r \eta_{r'} +  \sum_\alpha E^\alpha_{rr'}\h{\eta}_{r}\sigma^\alpha \eta_{r'}\right),
\end{equation*}
where the single occupancy constraint is implemented on average through chemical
potentials $\lambda_r$ tuned to enforce $\avg{\h{\eta}_r \eta_r}=1$. 

To gain insight into which
spin liquid may be favoured as we move away from the Heisenberg limit,
for each ansatz Hamiltonian $H(\chi,E)$ we compute the ground state $\ket{\psi(\chi,E)}$.
The energy $\epsilon(\chi,E) = \bra{\psi(\chi,E)} H_{\rm Pr}\ket{\psi(\chi,E)}$, where $H_{\rm Pr}$ is 
the full Pr Hamiltonian, is then minimized with respect to $\chi$ and $E$.
The phase diagram is shown in Fig. \ref{phase-diagram}, giving the state with lowest $\epsilon$
 as a function of $J_{\perp}/J_z$ and $J_{\pm\pm}/J_z$.  The monopole ansatz occupies large
region of the phase diagram around the Heisenberg point, with $E^z$ terms becoming finite at small $J_{\perp}$ and
the $E^{\pm}$ components remaining disfavoured throughout. The uniform state is fully
symmetric, with trivial PSG and does not become favoured until $J_{\pm\pm}$ is of order $\sim J_z/2$. The
ansatz has the simple form $\chi_{rr'} = \chi$, $E^+_{rr'} = \cb{\gamma}_{rr'}E$ and $E^z_{rr'}=0$. We show the
dispersion of this state when $E \neq 0$ and $\chi=E/2$ in Fig. \ref{uniform-bands-1}.
Note the lack of doubly degenerate bands, despite the presence of both
time-reversal and inversion symmetry, due to these pseudo-spinons being
non-Kramers.

\emph{Broken Symmetries:} We now consider the full Hamiltonian $H = H_{\rm Ir} + H_{\rm hyb}+H_{\rm Pr}$, adding in terms describing
a uniform spin liquid on the Pr.
When the $\Phi_5$ boson condenses the $U(1)\times U(1)$ gauge symmetry
of the decoupled electron and spin-liquid system is broken to a single $U(1)$\cite{senthil2003fractionalized,coleman2007heavy},
given by the transformation $\eta \rightarrow e^{i\theta} \eta$ 
and $c \rightarrow e^{i\theta} c$. This breaking of the relative gauge symmetry results in a Meissner-like
effect, with a mass term pinning the emergent and physical gauge fields together. This pinning
manifests in the acquisition of electric charge by the pseudo-spinon $\eta$ and allowing the $\eta$
pseudo-spinons to contribution directly to the Fermi sea as well as electromagnetic properties of the
system\cite{coleman2005transport}. A more general problem, which can be accessed by considering
further intermediate $4f^1$ and $4f^3$ channels, is an arbitrary hybridization
\begin{equation}
  H_{\rm hyb} \sim \sum_{rr'} \sum_{\alpha\beta} V^{\alpha\beta}_{rr'} \h{c}_{r\alpha} \eta_{r'\beta} +{\rm h.c}
\end{equation}
Any choice of this $V_{rr'}$, when both $H_{\rm Ir}$ and $H_{\rm Pr}$ are present, will result in not
only a breaking of time-reversal 
but in addition a breaking of at least one of the spatial symmetries. This is due to an incompatibility
between the gauge structures of the Ir and Pr sublattices. 
For all operations $S$ in $Fd\cb{3}m$, a symmetric hybridization
must have
\begin{equation}
  \label{symmetry-hybrid}
  V_{rr'} = z_{S,r} e^{i \theta_S} L_S V_{S^{-1}(r),S^{-1}(r')} \h{M}_S 
\end{equation}
for some choice of phases $e^{i\theta_S}$. 
Since the symmetries in the local axes
form a group, for any operations $S$ and $S'$ the action of $SS'$ must be
equivalent to the action of $S'$ followed by $S$. 
The local rotations satisfy $L_{SS'}=L_SL_{S'}$, so an equation relating 
$z_{SS',r}$ to $z_{S,r}$ and $z_{S',r}$ can be obtained.
Explicitly, this is given by\cite{wen2002quantum}
\begin{equation}
z_{SS',r} = \eta_{S,S'} z_{S,r} z_{S',S^{-1}(r)}  
\end{equation}
with $\eta_{S,S'} = \pm 1$. When
combined with Eq. \ref{symmetry-hybrid}
this consistency condition entails that $\eta_{S,S'}$ be gauge equivalent
to $1$. For the PSG of $z_{S,r}$ is false, and so this only
satisfied by some subgroup of $Fd\cb{3}m$, breaking some of the symmetry.

The specific form shown in Eq. \ref{hyb1}
motivated by the Anderson limit, breaks all spatial symmetries except for
inversion and a single $C_3$ axis. In the gauge used throughout the paper
this is the $[111]$ axis.
As shown in Fig. \ref{fluxes},
if spin dependence is ignored, then we can understand the gauge structure in a qualitative fashion as
a flux of $\pi/2$ exiting each tetrahedral face of the Ir sublattice. With the uniform
spin liquid on the Pr and $V_{rr'}$ on the Pr-Ir bonds chosen as in Eq. \ref{hyb1}, the flux is trapped
in this truncated tetrahedron. Since the flux is not exiting,
it must recombine into $2\pi$ flux somewhere within the volume. 
We have arranged it to preserve one of the $C_3$ axes.
When the Pr bonds are not present, this flux can cancel inside the 
remaining tetrahedra and thus form a symmetric state.
In the presence of Pr-Ir bonds, a flux passes through the plaquettes 
between the Pr and Ir breaking the symmetries.
\begin{figure}[tp]
\begin{subfigure}[b]{0.6\columnwidth} 
\begin{subfigure}[b]{\textwidth} 
  \includegraphics[width=0.8\textwidth]{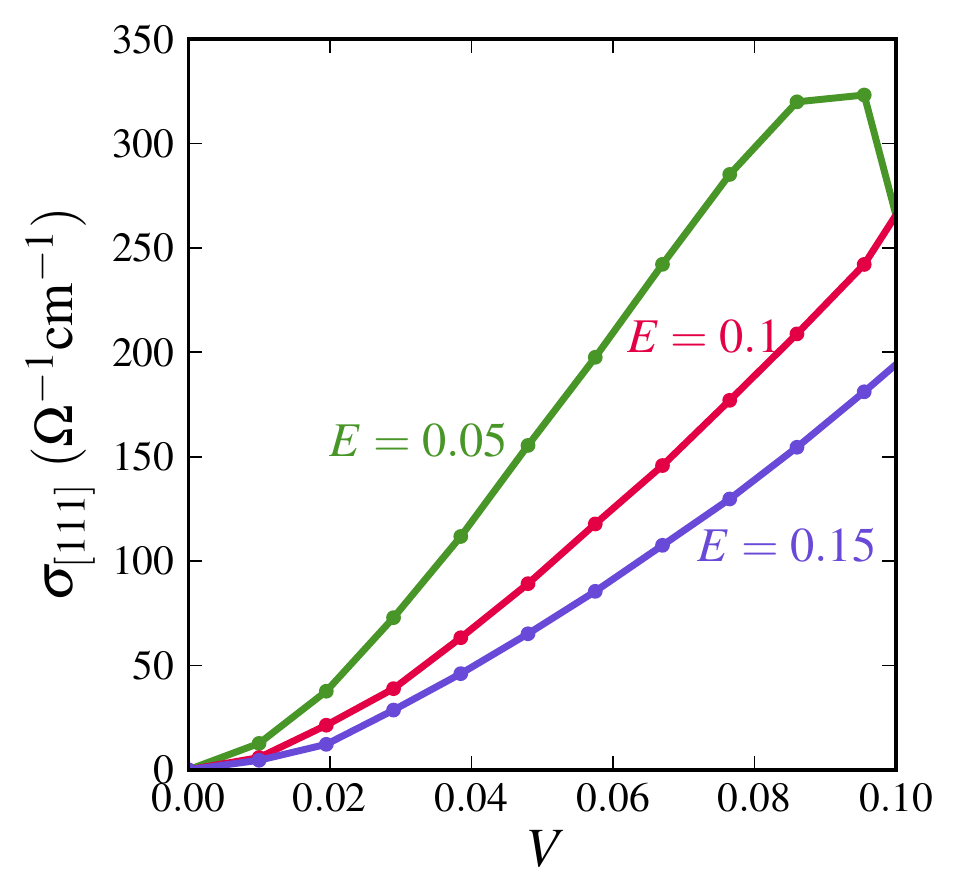}
  \caption{\label{results-sigma}AHE coefficient}
\end{subfigure} 
\begin{subfigure}[b]{\textwidth} 
  \includegraphics[width=0.8\textwidth]{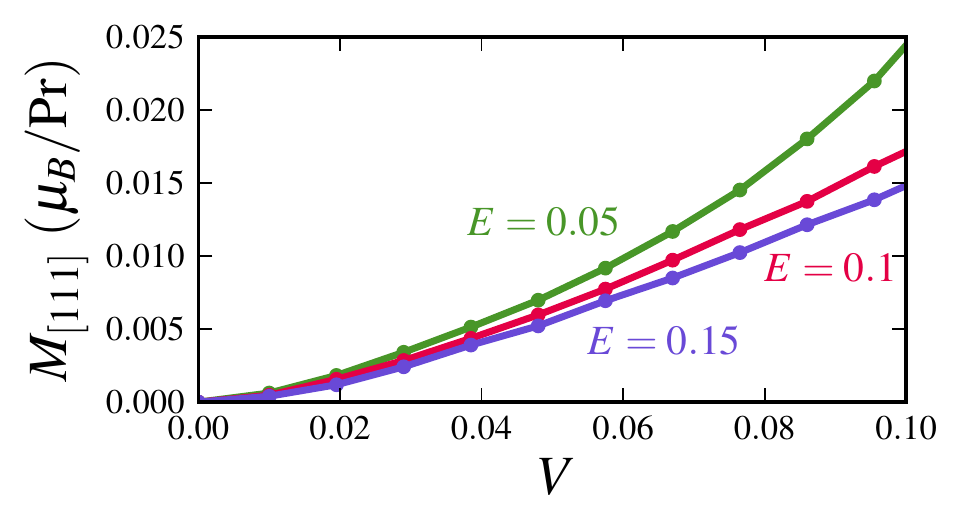}
  \caption{\label{results-m}Net magnetization}
\end{subfigure}
\end{subfigure}
\begin{subfigure}[b]{0.35\columnwidth}
\begin{subfigure}[b]{0.8\columnwidth} 
  \includegraphics[width=\textwidth]{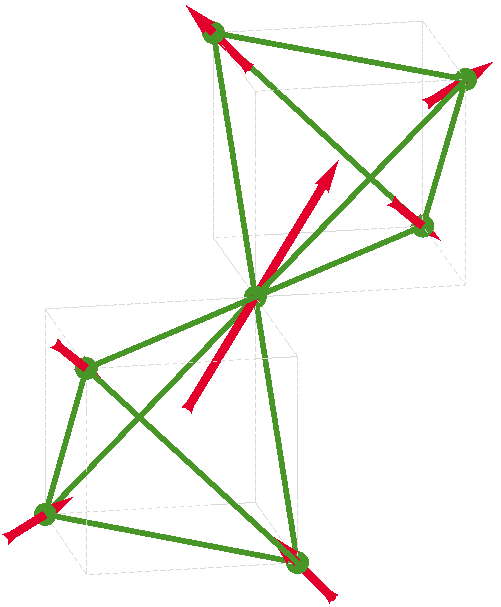}
  \caption{\label{moments-pr}Pr moments}
\end{subfigure}
\begin{subfigure}[b]{0.8\columnwidth} 
  \includegraphics[width=\textwidth]{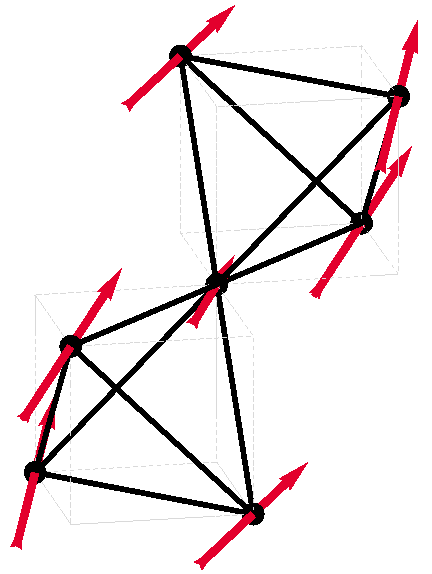}
  \caption{\label{moments-ir}Ir moments}
\end{subfigure}
\end{subfigure}
\caption{\label{results}\small
(a-b) The AHE and net magnetization along the $[111]$ direction for several values of $E$
as a function of $V$, with $t_2/t_1$ fixed at $0.1$ and $\chi/E = 0.5$.
(c-d) The pattern of local magnetic moments on the Ir and magnetic and quadrupolar
moments on the Pr. 
}
\end{figure}
With the only remaining symmetries are inversion
and single $C_3$ axis, the system is sufficiently asymmetric such that $\vec{\sigma}_A$
is allowed oriented along the $[111]$ direction. Further magnetic, charge
and quadrupolar orderings are generically induced, subject only to this fairly
permissive $C_3$ symmetry and inversion.

\emph{Physical consequences:} To explore the effects of the spin liquid parameters and hybridization we fix $t_1=1$, $t_2=0.1t_1$ and $\chi=E/2$
and vary $E$ and $V$. This assumes that $J_{\pm\pm}/J_z$ is sufficiently large so that a uniform spin liquid is stabilized.
Calculations of magnetization and AHE coefficients are
shown in Fig. \ref{results-sigma} and Fig. \ref{results-m}. The magnetization shows
the net magnetic moment per Pr atom, oriented along the $[111]$, with contributions from both Pr and Ir sublattices
(as shown in Fig. \ref{moments-pr} and \ref{moments-ir})
using $g$ factors of $g_{\rm Pr} \sim 6.0$ and $g_{\rm Ir} \sim 2.0$.
The anomalous Hall vector $\vec{\sigma}_A$ is computed using the Kubo formula\cite{nagaosa2010anomalous}, where
the pseudo-spinons contribute as electrons to the current operators in the condensed phase.

The large AHE with small magnetic moments is in
qualitative agreement with the properties of the intermediate phase of Pr$_2$Ir$_2$O$_7$. Here both the AHE and
magnetic moment are considerably larger than observed experimentally. This discrepancy can be explained
if the domains of the ordered phase are not fully aligned by the hysteresis process, then the observed
AHE and magnetization would represent residual contributions from the partially aligned domains.
At the mean field level one expects that the transition into the hybridized phase should show
a jump in the specific heat.  Since the order parameter
can take 8 directions along the $[111]$ axes one expects the critical theory to be described by an $O(3)$
type model, leading to a cusp at the transition. The effects disorder or Pr-Ir substitution can potentially
smooth this cusp into the broad peak seen in experiments\cite{machida2009time} once background
contributions have been subtracted. The onset of hybridization between the pseudo-spinons 
and the electron alters the electronic band structure. How this manifests at 
the transition depends on the gauge fluctuations, as the binding of electric charge to
the pseudo-spinons softens as one approaches the critical point.

An essential feature of our proposal is the lack of large on-site moments. All
induced orderings, such as the magnetic, quadrupolar and charge modulations are
small, only appearing at fractions $\sim 10^{-2}-10^{-3}$ of their
saturated values. Distinct from scenarios with
large moments that approximately cancel, leaving a small net moment. To
distinguish these experimentally, nuclear magnetic resonance (NMR) on the oxygens
is promising. In the crystal structure the oxygens lie in two inequivalent Wyckoff positions:
the $8a$ position, which has tetrahedral symmetry and the $48f$ position which is
in an asymmetric location. If one can account for small net moment through a 
cancellation of large local moments, then one may expect the net field
at the symmetric site $8a$ to be small, but generically asymmetric site $48f$ 
should be affected by a net field from moments of order $\sim \mu_B$.
In this scenario assuming dipolar
fields from the moments acting at the oxygen sites, one then expects the
effect at the $48f$ site to be several orders of magnitude larger than that
of the $8a$ site. Our proposal predicts a significantly different result, with
the small local moments producing only small fields of order $\sim 0.1-1G$ at both
oxygen sites.

\emph{Conclusion:} We have proposed a mechanism for symmetry breaking when conduction
electrons hybridize with a quantum spin liquid.
Applied to Pr$_2$Ir$_2$O$_7$ we found that the hybridization of two subsystems ($f$ and $d$ electrons) results in a
chiral nematic metal with broken time reversal and spatial symmetries, exhibiting an
anomalous Hall effect without a sizeable magnetic moment.
This mechanism could potentially manifest in a wide range of heavy fermion materials on geometrically frustrated lattices.

\emph{Acknowledgments:} We would like to thank S.B. Lee, Y.B. Kim, L. Balents, S. Bhattacharjee
and A. Paramekanti for helpful discussions.
This work was supported by the NSERC of Canada.

\bibliography{draft}

\begin{thebibliography}{36}%
\makeatletter
\providecommand \@ifxundefined [1]{%
 \@ifx{#1\undefined}
}%
\providecommand \@ifnum [1]{%
 \ifnum #1\expandafter \@firstoftwo
 \else \expandafter \@secondoftwo
 \fi
}%
\providecommand \@ifx [1]{%
 \ifx #1\expandafter \@firstoftwo
 \else \expandafter \@secondoftwo
 \fi
}%
\providecommand \natexlab [1]{#1}%
\providecommand \enquote  [1]{``#1''}%
\providecommand \bibnamefont  [1]{#1}%
\providecommand \bibfnamefont [1]{#1}%
\providecommand \citenamefont [1]{#1}%
\providecommand \href@noop [0]{\@secondoftwo}%
\providecommand \href [0]{\begingroup \@sanitize@url \@href}%
\providecommand \@href[1]{\@@startlink{#1}\@@href}%
\providecommand \@@href[1]{\endgroup#1\@@endlink}%
\providecommand \@sanitize@url [0]{\catcode `\\12\catcode `\$12\catcode
  `\&12\catcode `\#12\catcode `\^12\catcode `\_12\catcode `\%12\relax}%
\providecommand \@@startlink[1]{}%
\providecommand \@@endlink[0]{}%
\providecommand \url  [0]{\begingroup\@sanitize@url \@url }%
\providecommand \@url [1]{\endgroup\@href {#1}{\urlprefix }}%
\providecommand \urlprefix  [0]{URL }%
\providecommand \Eprint [0]{\href }%
\providecommand \doibase [0]{http://dx.doi.org/}%
\providecommand \selectlanguage [0]{\@gobble}%
\providecommand \bibinfo  [0]{\@secondoftwo}%
\providecommand \bibfield  [0]{\@secondoftwo}%
\providecommand \translation [1]{[#1]}%
\providecommand \BibitemOpen [0]{}%
\providecommand \bibitemStop [0]{}%
\providecommand \bibitemNoStop [0]{.\EOS\space}%
\providecommand \EOS [0]{\spacefactor3000\relax}%
\providecommand \BibitemShut  [1]{\csname bibitem#1\endcsname}%
\let\auto@bib@innerbib\@empty
\bibitem [{\citenamefont {Kondo}(1964)}]{kondo1964resistance}%
  \BibitemOpen
  \bibfield  {author} {\bibinfo {author} {\bibfnamefont {J.}~\bibnamefont
  {Kondo}},\ }\href@noop {} {\bibfield  {journal} {\bibinfo  {journal}
  {Progress of theoretical physics}\ }\textbf {\bibinfo {volume} {32}},\
  \bibinfo {pages} {37} (\bibinfo {year} {1964})}\BibitemShut {NoStop}%
\bibitem [{\citenamefont {Wilson}(1975)}]{wilson1975renormalization}%
  \BibitemOpen
  \bibfield  {author} {\bibinfo {author} {\bibfnamefont {K.~G.}\ \bibnamefont
  {Wilson}},\ }\href@noop {} {\bibfield  {journal} {\bibinfo  {journal}
  {Reviews of Modern Physics}\ }\textbf {\bibinfo {volume} {47}},\ \bibinfo
  {pages} {773} (\bibinfo {year} {1975})}\BibitemShut {NoStop}%
\bibitem [{\citenamefont {Stewart}(1984)}]{stewart1984heavy}%
  \BibitemOpen
  \bibfield  {author} {\bibinfo {author} {\bibfnamefont {G.}~\bibnamefont
  {Stewart}},\ }\href@noop {} {\bibfield  {journal} {\bibinfo  {journal}
  {Reviews of Modern Physics}\ }\textbf {\bibinfo {volume} {56}},\ \bibinfo
  {pages} {755} (\bibinfo {year} {1984})}\BibitemShut {NoStop}%
\bibitem [{\citenamefont {Doniach}(1977)}]{doniach1977kondo}%
  \BibitemOpen
  \bibfield  {author} {\bibinfo {author} {\bibfnamefont {S.}~\bibnamefont
  {Doniach}},\ }\href@noop {} {\bibfield  {journal} {\bibinfo  {journal}
  {Physica B+ C}\ }\textbf {\bibinfo {volume} {91}},\ \bibinfo {pages} {231}
  (\bibinfo {year} {1977})}\BibitemShut {NoStop}%
\bibitem [{\citenamefont {Hewson}(1997)}]{hewson1997kondo}%
  \BibitemOpen
  \bibfield  {author} {\bibinfo {author} {\bibfnamefont {A.~C.}\ \bibnamefont
  {Hewson}},\ }\href@noop {} {\emph {\bibinfo {title} {The Kondo problem to
  heavy fermions}}},\ Vol.~\bibinfo {volume} {2}\ (\bibinfo  {publisher}
  {Cambridge university press},\ \bibinfo {year} {1997})\BibitemShut {NoStop}%
\bibitem [{\citenamefont {Nagaosa}\ \emph {et~al.}(2010)\citenamefont
  {Nagaosa}, \citenamefont {Sinova}, \citenamefont {Onoda}, \citenamefont
  {MacDonald},\ and\ \citenamefont {Ong}}]{nagaosa2010anomalous}%
  \BibitemOpen
  \bibfield  {author} {\bibinfo {author} {\bibfnamefont {N.}~\bibnamefont
  {Nagaosa}}, \bibinfo {author} {\bibfnamefont {J.}~\bibnamefont {Sinova}},
  \bibinfo {author} {\bibfnamefont {S.}~\bibnamefont {Onoda}}, \bibinfo
  {author} {\bibfnamefont {A.}~\bibnamefont {MacDonald}}, \ and\ \bibinfo
  {author} {\bibfnamefont {N.}~\bibnamefont {Ong}},\ }\href@noop {} {\bibfield
  {journal} {\bibinfo  {journal} {Reviews of Modern Physics}\ }\textbf
  {\bibinfo {volume} {82}},\ \bibinfo {pages} {1539} (\bibinfo {year}
  {2010})}\BibitemShut {NoStop}%
\bibitem [{\citenamefont {Si}(2006)}]{si2006global}%
  \BibitemOpen
  \bibfield  {author} {\bibinfo {author} {\bibfnamefont {Q.}~\bibnamefont
  {Si}},\ }\href@noop {} {\bibfield  {journal} {\bibinfo  {journal} {Physica B:
  Condensed Matter}\ }\textbf {\bibinfo {volume} {378}},\ \bibinfo {pages} {23}
  (\bibinfo {year} {2006})}\BibitemShut {NoStop}%
\bibitem [{\citenamefont {Senthil}\ \emph {et~al.}(2003)\citenamefont
  {Senthil}, \citenamefont {Sachdev},\ and\ \citenamefont
  {Vojta}}]{senthil2003fractionalized}%
  \BibitemOpen
  \bibfield  {author} {\bibinfo {author} {\bibfnamefont {T.}~\bibnamefont
  {Senthil}}, \bibinfo {author} {\bibfnamefont {S.}~\bibnamefont {Sachdev}}, \
  and\ \bibinfo {author} {\bibfnamefont {M.}~\bibnamefont {Vojta}},\
  }\href@noop {} {\bibfield  {journal} {\bibinfo  {journal} {Physical review
  letters}\ }\textbf {\bibinfo {volume} {90}},\ \bibinfo {pages} {216403}
  (\bibinfo {year} {2003})}\BibitemShut {NoStop}%
\bibitem [{\citenamefont {Senthil}\ \emph {et~al.}(2004)\citenamefont
  {Senthil}, \citenamefont {Vojta},\ and\ \citenamefont
  {Sachdev}}]{senthil2004weak}%
  \BibitemOpen
  \bibfield  {author} {\bibinfo {author} {\bibfnamefont {T.}~\bibnamefont
  {Senthil}}, \bibinfo {author} {\bibfnamefont {M.}~\bibnamefont {Vojta}}, \
  and\ \bibinfo {author} {\bibfnamefont {S.}~\bibnamefont {Sachdev}},\
  }\href@noop {} {\bibfield  {journal} {\bibinfo  {journal} {Physical Review
  B}\ }\textbf {\bibinfo {volume} {69}},\ \bibinfo {pages} {035111} (\bibinfo
  {year} {2004})}\BibitemShut {NoStop}%
\bibitem [{\citenamefont {Ghaemi}\ and\ \citenamefont
  {Senthil}(2007)}]{ghaemi2007higher}%
  \BibitemOpen
  \bibfield  {author} {\bibinfo {author} {\bibfnamefont {P.}~\bibnamefont
  {Ghaemi}}\ and\ \bibinfo {author} {\bibfnamefont {T.}~\bibnamefont
  {Senthil}},\ }\href@noop {} {\bibfield  {journal} {\bibinfo  {journal}
  {Physical Review B}\ }\textbf {\bibinfo {volume} {75}},\ \bibinfo {pages}
  {144412} (\bibinfo {year} {2007})}\BibitemShut {NoStop}%
\bibitem [{\citenamefont {Coleman}\ and\ \citenamefont
  {Nevidomskyy}(2010)}]{coleman2010frustration}%
  \BibitemOpen
  \bibfield  {author} {\bibinfo {author} {\bibfnamefont {P.}~\bibnamefont
  {Coleman}}\ and\ \bibinfo {author} {\bibfnamefont {A.~H.}\ \bibnamefont
  {Nevidomskyy}},\ }\href@noop {} {\bibfield  {journal} {\bibinfo  {journal}
  {Journal of Low Temperature Physics}\ }\textbf {\bibinfo {volume} {161}},\
  \bibinfo {pages} {182} (\bibinfo {year} {2010})}\BibitemShut {NoStop}%
\bibitem [{\citenamefont {MacLaughlin}\ \emph {et~al.}(2009)\citenamefont
  {MacLaughlin}, \citenamefont {Ohta}, \citenamefont {Machida}, \citenamefont
  {Nakatsuji}, \citenamefont {Luke}, \citenamefont {Ishida}, \citenamefont
  {Heffner}, \citenamefont {Shu},\ and\ \citenamefont
  {Bernal}}]{maclaughlin2009weak}%
  \BibitemOpen
  \bibfield  {author} {\bibinfo {author} {\bibfnamefont {D.}~\bibnamefont
  {MacLaughlin}}, \bibinfo {author} {\bibfnamefont {Y.}~\bibnamefont {Ohta}},
  \bibinfo {author} {\bibfnamefont {Y.}~\bibnamefont {Machida}}, \bibinfo
  {author} {\bibfnamefont {S.}~\bibnamefont {Nakatsuji}}, \bibinfo {author}
  {\bibfnamefont {G.}~\bibnamefont {Luke}}, \bibinfo {author} {\bibfnamefont
  {K.}~\bibnamefont {Ishida}}, \bibinfo {author} {\bibfnamefont
  {R.}~\bibnamefont {Heffner}}, \bibinfo {author} {\bibfnamefont
  {L.}~\bibnamefont {Shu}}, \ and\ \bibinfo {author} {\bibfnamefont
  {O.}~\bibnamefont {Bernal}},\ }\href@noop {} {\bibfield  {journal} {\bibinfo
  {journal} {Physica B: Condensed Matter}\ }\textbf {\bibinfo {volume} {404}},\
  \bibinfo {pages} {667} (\bibinfo {year} {2009})}\BibitemShut {NoStop}%
\bibitem [{\citenamefont {Nakatsuji}\ \emph {et~al.}(2006)\citenamefont
  {Nakatsuji}, \citenamefont {Machida}, \citenamefont {Maeno}, \citenamefont
  {Tayama}, \citenamefont {Sakakibara}, \citenamefont {Duijn}, \citenamefont
  {Balicas}, \citenamefont {Millican}, \citenamefont {Macaluso},\ and\
  \citenamefont {Chan}}]{nakatsuji2006metallic}%
  \BibitemOpen
  \bibfield  {author} {\bibinfo {author} {\bibfnamefont {S.}~\bibnamefont
  {Nakatsuji}}, \bibinfo {author} {\bibfnamefont {Y.}~\bibnamefont {Machida}},
  \bibinfo {author} {\bibfnamefont {Y.}~\bibnamefont {Maeno}}, \bibinfo
  {author} {\bibfnamefont {T.}~\bibnamefont {Tayama}}, \bibinfo {author}
  {\bibfnamefont {T.}~\bibnamefont {Sakakibara}}, \bibinfo {author}
  {\bibfnamefont {J.~v.}\ \bibnamefont {Duijn}}, \bibinfo {author}
  {\bibfnamefont {L.}~\bibnamefont {Balicas}}, \bibinfo {author} {\bibfnamefont
  {J.}~\bibnamefont {Millican}}, \bibinfo {author} {\bibfnamefont
  {R.}~\bibnamefont {Macaluso}}, \ and\ \bibinfo {author} {\bibfnamefont
  {J.~Y.}\ \bibnamefont {Chan}},\ }\href@noop {} {\bibfield  {journal}
  {\bibinfo  {journal} {Physical review letters}\ }\textbf {\bibinfo {volume}
  {96}},\ \bibinfo {pages} {087204} (\bibinfo {year} {2006})}\BibitemShut
  {NoStop}%
\bibitem [{\citenamefont {Machida}\ \emph {et~al.}(2007)\citenamefont
  {Machida}, \citenamefont {Nakatsuji}, \citenamefont {Maeno}, \citenamefont
  {Tayama}, \citenamefont {Sakakibara},\ and\ \citenamefont
  {Onoda}}]{machida2007unconventional}%
  \BibitemOpen
  \bibfield  {author} {\bibinfo {author} {\bibfnamefont {Y.}~\bibnamefont
  {Machida}}, \bibinfo {author} {\bibfnamefont {S.}~\bibnamefont {Nakatsuji}},
  \bibinfo {author} {\bibfnamefont {Y.}~\bibnamefont {Maeno}}, \bibinfo
  {author} {\bibfnamefont {T.}~\bibnamefont {Tayama}}, \bibinfo {author}
  {\bibfnamefont {T.}~\bibnamefont {Sakakibara}}, \ and\ \bibinfo {author}
  {\bibfnamefont {S.}~\bibnamefont {Onoda}},\ }\href@noop {} {\bibfield
  {journal} {\bibinfo  {journal} {Physical review letters}\ }\textbf {\bibinfo
  {volume} {98}},\ \bibinfo {pages} {057203} (\bibinfo {year}
  {2007})}\BibitemShut {NoStop}%
\bibitem [{\citenamefont {Zhou}\ \emph {et~al.}(2008)\citenamefont {Zhou},
  \citenamefont {Wiebe}, \citenamefont {Janik}, \citenamefont {Balicas},
  \citenamefont {Yo}, \citenamefont {Qiu}, \citenamefont {Copley},\ and\
  \citenamefont {Gardner}}]{zhou2008dynamic}%
  \BibitemOpen
  \bibfield  {author} {\bibinfo {author} {\bibfnamefont {H.}~\bibnamefont
  {Zhou}}, \bibinfo {author} {\bibfnamefont {C.}~\bibnamefont {Wiebe}},
  \bibinfo {author} {\bibfnamefont {J.}~\bibnamefont {Janik}}, \bibinfo
  {author} {\bibfnamefont {L.}~\bibnamefont {Balicas}}, \bibinfo {author}
  {\bibfnamefont {Y.}~\bibnamefont {Yo}}, \bibinfo {author} {\bibfnamefont
  {Y.}~\bibnamefont {Qiu}}, \bibinfo {author} {\bibfnamefont {J.}~\bibnamefont
  {Copley}}, \ and\ \bibinfo {author} {\bibfnamefont {J.}~\bibnamefont
  {Gardner}},\ }\href@noop {} {\bibfield  {journal} {\bibinfo  {journal}
  {Physical review letters}\ }\textbf {\bibinfo {volume} {101}},\ \bibinfo
  {pages} {227204} (\bibinfo {year} {2008})}\BibitemShut {NoStop}%
\bibitem [{\citenamefont {Machida}\ \emph {et~al.}(2009)\citenamefont
  {Machida}, \citenamefont {Nakatsuji}, \citenamefont {Onoda}, \citenamefont
  {Tayama},\ and\ \citenamefont {Sakakibara}}]{machida2009time}%
  \BibitemOpen
  \bibfield  {author} {\bibinfo {author} {\bibfnamefont {Y.}~\bibnamefont
  {Machida}}, \bibinfo {author} {\bibfnamefont {S.}~\bibnamefont {Nakatsuji}},
  \bibinfo {author} {\bibfnamefont {S.}~\bibnamefont {Onoda}}, \bibinfo
  {author} {\bibfnamefont {T.}~\bibnamefont {Tayama}}, \ and\ \bibinfo {author}
  {\bibfnamefont {T.}~\bibnamefont {Sakakibara}},\ }\href@noop {} {\bibfield
  {journal} {\bibinfo  {journal} {Nature}\ }\textbf {\bibinfo {volume} {463}},\
  \bibinfo {pages} {210} (\bibinfo {year} {2009})}\BibitemShut {NoStop}%
\bibitem [{\citenamefont {Udagawa}\ and\ \citenamefont
  {Moessner}(2012)}]{udagawa2012anomalous}%
  \BibitemOpen
  \bibfield  {author} {\bibinfo {author} {\bibfnamefont {M.}~\bibnamefont
  {Udagawa}}\ and\ \bibinfo {author} {\bibfnamefont {R.}~\bibnamefont
  {Moessner}},\ }\href@noop {} {\bibfield  {journal} {\bibinfo  {journal}
  {arXiv preprint arXiv:1212.0293}\ } (\bibinfo {year} {2012})}\BibitemShut
  {NoStop}%
\bibitem [{\citenamefont {Moon}\ \emph {et~al.}(2012)\citenamefont {Moon},
  \citenamefont {Xu}, \citenamefont {Kim},\ and\ \citenamefont
  {Balents}}]{moon2012non}%
  \BibitemOpen
  \bibfield  {author} {\bibinfo {author} {\bibfnamefont {E.-G.}\ \bibnamefont
  {Moon}}, \bibinfo {author} {\bibfnamefont {C.}~\bibnamefont {Xu}}, \bibinfo
  {author} {\bibfnamefont {Y.~B.}\ \bibnamefont {Kim}}, \ and\ \bibinfo
  {author} {\bibfnamefont {L.}~\bibnamefont {Balents}},\ }\href@noop {}
  {\bibfield  {journal} {\bibinfo  {journal} {arXiv preprint arXiv:1212.1168}\
  } (\bibinfo {year} {2012})}\BibitemShut {NoStop}%
\bibitem [{\citenamefont {Flint}\ and\ \citenamefont
  {Senthil}(2013)}]{flint2013chiral}%
  \BibitemOpen
  \bibfield  {author} {\bibinfo {author} {\bibfnamefont {R.}~\bibnamefont
  {Flint}}\ and\ \bibinfo {author} {\bibfnamefont {T.}~\bibnamefont
  {Senthil}},\ } {\bibfield
  {journal} {\bibinfo  {journal} {Phys. Rev. B}\ }\textbf {\bibinfo {volume}
  {87}},\ \bibinfo {pages} {125147} (\bibinfo {year} {2013})}\BibitemShut
  {NoStop}%
\bibitem [{\citenamefont {Chen}\ and\ \citenamefont
  {Hermele}(2012)}]{chen2012magnetic}%
  \BibitemOpen
  \bibfield  {author} {\bibinfo {author} {\bibfnamefont {G.}~\bibnamefont
  {Chen}}\ and\ \bibinfo {author} {\bibfnamefont {M.}~\bibnamefont {Hermele}},\
  }\href@noop {} {\bibfield  {journal} {\bibinfo  {journal} {Physical Review
  B}\ }\textbf {\bibinfo {volume} {86}},\ \bibinfo {pages} {235129} (\bibinfo
  {year} {2012})}\BibitemShut {NoStop}%
\bibitem [{\citenamefont {Lee}\ \emph {et~al.}(2013)\citenamefont {Lee},
  \citenamefont {Paramekanti},\ and\ \citenamefont {Kim}}]{lee2013rkky}%
  \BibitemOpen
  \bibfield  {author} {\bibinfo {author} {\bibfnamefont {S.}~\bibnamefont
  {Lee}}, \bibinfo {author} {\bibfnamefont {A.}~\bibnamefont {Paramekanti}}, \
  and\ \bibinfo {author} {\bibfnamefont {Y.~B.}\ \bibnamefont {Kim}},\
  }\href@noop {} {\bibfield  {journal} {\bibinfo  {journal} {arXiv preprint
  arXiv:1305.0827}\ } (\bibinfo {year} {2013})}\BibitemShut {NoStop}%
\bibitem [{\citenamefont {Kim}\ \emph {et~al.}(2008)\citenamefont {Kim},
  \citenamefont {Jin}, \citenamefont {Moon}, \citenamefont {Kim}, \citenamefont
  {Park}, \citenamefont {Leem}, \citenamefont {Yu}, \citenamefont {Noh},
  \citenamefont {Kim}, \citenamefont {Oh} \emph {et~al.}}]{kim2008novel}%
  \BibitemOpen
  \bibfield  {author} {\bibinfo {author} {\bibfnamefont {B.}~\bibnamefont
  {Kim}}, \bibinfo {author} {\bibfnamefont {H.}~\bibnamefont {Jin}}, \bibinfo
  {author} {\bibfnamefont {S.}~\bibnamefont {Moon}}, \bibinfo {author}
  {\bibfnamefont {J.-Y.}\ \bibnamefont {Kim}}, \bibinfo {author} {\bibfnamefont
  {B.-G.}\ \bibnamefont {Park}}, \bibinfo {author} {\bibfnamefont
  {C.}~\bibnamefont {Leem}}, \bibinfo {author} {\bibfnamefont {J.}~\bibnamefont
  {Yu}}, \bibinfo {author} {\bibfnamefont {T.}~\bibnamefont {Noh}}, \bibinfo
  {author} {\bibfnamefont {C.}~\bibnamefont {Kim}}, \bibinfo {author}
  {\bibfnamefont {S.-J.}\ \bibnamefont {Oh}},  \emph {et~al.},\ }\href@noop {}
  {\bibfield  {journal} {\bibinfo  {journal} {Physical Review Letters}\
  }\textbf {\bibinfo {volume} {101}},\ \bibinfo {pages} {076402} (\bibinfo
  {year} {2008})}\BibitemShut {NoStop}%
\bibitem [{\citenamefont {Burnell}\ \emph {et~al.}(2009)\citenamefont
  {Burnell}, \citenamefont {Chakravarty},\ and\ \citenamefont
  {Sondhi}}]{burnell2009monopole}%
  \BibitemOpen
  \bibfield  {author} {\bibinfo {author} {\bibfnamefont {F.}~\bibnamefont
  {Burnell}}, \bibinfo {author} {\bibfnamefont {S.}~\bibnamefont
  {Chakravarty}}, \ and\ \bibinfo {author} {\bibfnamefont {S.}~\bibnamefont
  {Sondhi}},\ }\href@noop {} {\bibfield  {journal} {\bibinfo  {journal}
  {Physical Review B}\ }\textbf {\bibinfo {volume} {79}},\ \bibinfo {pages}
  {144432} (\bibinfo {year} {2009})}\BibitemShut {NoStop}%
\bibitem [{\citenamefont {Kurita}\ \emph {et~al.}(2011)\citenamefont {Kurita},
  \citenamefont {Yamaji},\ and\ \citenamefont {Imada}}]{kurita2011topological}%
  \BibitemOpen
  \bibfield  {author} {\bibinfo {author} {\bibfnamefont {M.}~\bibnamefont
  {Kurita}}, \bibinfo {author} {\bibfnamefont {Y.}~\bibnamefont {Yamaji}}, \
  and\ \bibinfo {author} {\bibfnamefont {M.}~\bibnamefont {Imada}},\ } {\bibfield  {journal} {\bibinfo  {journal}
  {Journal of the Physical Society of Japan}\ }\textbf {\bibinfo {volume}
  {80}},\ \bibinfo {pages} {044708} (\bibinfo {year} {2011})}\BibitemShut
  {NoStop}%
\bibitem [{\citenamefont {Witczak-Krempa}\ \emph {et~al.}(2013)\citenamefont
  {Witczak-Krempa}, \citenamefont {Go},\ and\ \citenamefont
  {Kim}}]{krempa2013pyrochlore}%
  \BibitemOpen
  \bibfield  {author} {\bibinfo {author} {\bibfnamefont {W.}~\bibnamefont
  {Witczak-Krempa}}, \bibinfo {author} {\bibfnamefont {A.}~\bibnamefont {Go}},
  \ and\ \bibinfo {author} {\bibfnamefont {Y.~B.}\ \bibnamefont {Kim}},\ } {\bibfield  {journal} {\bibinfo
  {journal} {Phys. Rev. B}\ }\textbf {\bibinfo {volume} {87}},\ \bibinfo
  {pages} {155101} (\bibinfo {year} {2013})}\BibitemShut {NoStop}%
\bibitem [{\citenamefont {Machida}\ \emph {et~al.}(2005)\citenamefont
  {Machida}, \citenamefont {Nakatsuji}, \citenamefont {Tonomura}, \citenamefont
  {Tayama}, \citenamefont {Sakakibara}, \citenamefont {Van~Duijn},
  \citenamefont {Broholm},\ and\ \citenamefont
  {Maeno}}]{machida2005crystalline}%
  \BibitemOpen
  \bibfield  {author} {\bibinfo {author} {\bibfnamefont {Y.}~\bibnamefont
  {Machida}}, \bibinfo {author} {\bibfnamefont {S.}~\bibnamefont {Nakatsuji}},
  \bibinfo {author} {\bibfnamefont {H.}~\bibnamefont {Tonomura}}, \bibinfo
  {author} {\bibfnamefont {T.}~\bibnamefont {Tayama}}, \bibinfo {author}
  {\bibfnamefont {T.}~\bibnamefont {Sakakibara}}, \bibinfo {author}
  {\bibfnamefont {J.}~\bibnamefont {Van~Duijn}}, \bibinfo {author}
  {\bibfnamefont {C.}~\bibnamefont {Broholm}}, \ and\ \bibinfo {author}
  {\bibfnamefont {Y.}~\bibnamefont {Maeno}},\ }\href@noop {} {\bibfield
  {journal} {\bibinfo  {journal} {Journal of Physics and Chemistry of Solids}\
  }\textbf {\bibinfo {volume} {66}},\ \bibinfo {pages} {1435} (\bibinfo {year}
  {2005})}\BibitemShut {NoStop}%
\bibitem [{\citenamefont {Onoda}\ and\ \citenamefont
  {Tanaka}(2010)}]{onoda2010quantum}%
  \BibitemOpen
  \bibfield  {author} {\bibinfo {author} {\bibfnamefont {S.}~\bibnamefont
  {Onoda}}\ and\ \bibinfo {author} {\bibfnamefont {Y.}~\bibnamefont {Tanaka}},\
  }\href@noop {} {\bibfield  {journal} {\bibinfo  {journal} {Physical review
  letters}\ }\textbf {\bibinfo {volume} {105}},\ \bibinfo {pages} {047201}
  (\bibinfo {year} {2010})}\BibitemShut {NoStop}%
\bibitem [{foo()}]{footnote1}%
  \BibitemOpen
  \href@noop {} {}\bibinfo {note} {The degeneracy of the non-Kramers doublet
  could be lifted, in principle, by a Jahn-Teller distortion of the surrounding
  oxygens. This would give rise to to onsite terms such as $\cc{Q} \tau^+ +Q
  \tau^-$. Due to the lack of evidence for any significant splitting from
  experiments, we ignore this term.}\BibitemShut {Stop}%
\bibitem [{\citenamefont {Onoda}\ and\ \citenamefont
  {Tanaka}(2011)}]{onoda2011quantum}%
  \BibitemOpen
  \bibfield  {author} {\bibinfo {author} {\bibfnamefont {S.}~\bibnamefont
  {Onoda}}\ and\ \bibinfo {author} {\bibfnamefont {Y.}~\bibnamefont {Tanaka}},\
  }\href@noop {} {\bibfield  {journal} {\bibinfo  {journal} {Physical Review
  B}\ }\textbf {\bibinfo {volume} {83}},\ \bibinfo {pages} {094411} (\bibinfo
  {year} {2011})}\BibitemShut {NoStop}%
\bibitem [{\citenamefont {Lee}\ \emph {et~al.}(2012)\citenamefont {Lee},
  \citenamefont {Onoda},\ and\ \citenamefont {Balents}}]{lee2012generic}%
  \BibitemOpen
  \bibfield  {author} {\bibinfo {author} {\bibfnamefont {S.}~\bibnamefont
  {Lee}}, \bibinfo {author} {\bibfnamefont {S.}~\bibnamefont {Onoda}}, \ and\
  \bibinfo {author} {\bibfnamefont {L.}~\bibnamefont {Balents}},\ }\href@noop
  {} {\bibfield  {journal} {\bibinfo  {journal} {Physical Review B}\ }\textbf
  {\bibinfo {volume} {86}},\ \bibinfo {pages} {104412} (\bibinfo {year}
  {2012})}\BibitemShut {NoStop}%
\bibitem [{\citenamefont {Bradley}\ and\ \citenamefont
  {Cracknell}(1972)}]{bradley1972mathematical}%
  \BibitemOpen
  \bibfield  {author} {\bibinfo {author} {\bibfnamefont {C.~J.}\ \bibnamefont
  {Bradley}}\ and\ \bibinfo {author} {\bibfnamefont {A.~P.}\ \bibnamefont
  {Cracknell}},\ }\href@noop {} {\emph {\bibinfo {title} {The mathematical
  theory of symmetry in solids: representation theory for point groups and
  space groups}}}\ (\bibinfo  {publisher} {Clarendon Press Oxford},\ \bibinfo
  {year} {1972})\BibitemShut {NoStop}%
\bibitem [{\citenamefont {Chandra}\ \emph {et~al.}(2013)\citenamefont
  {Chandra}, \citenamefont {Coleman},\ and\ \citenamefont
  {Flint}}]{chandra2013hastatic}%
  \BibitemOpen
  \bibfield  {author} {\bibinfo {author} {\bibfnamefont {P.}~\bibnamefont
  {Chandra}}, \bibinfo {author} {\bibfnamefont {P.}~\bibnamefont {Coleman}}, \
  and\ \bibinfo {author} {\bibfnamefont {R.}~\bibnamefont {Flint}},\
  }\href@noop {} {\bibfield  {journal} {\bibinfo  {journal} {Nature}\ }\textbf
  {\bibinfo {volume} {493}},\ \bibinfo {pages} {621} (\bibinfo {year}
  {2013})}\BibitemShut {NoStop}%
\bibitem [{\citenamefont {Kim}\ and\ \citenamefont
  {Han}(2008)}]{kim2008chiral}%
  \BibitemOpen
  \bibfield  {author} {\bibinfo {author} {\bibfnamefont {J.~H.}\ \bibnamefont
  {Kim}}\ and\ \bibinfo {author} {\bibfnamefont {J.~H.}\ \bibnamefont {Han}},\
  }\href@noop {} {\bibfield  {journal} {\bibinfo  {journal} {Physical Review
  B}\ }\textbf {\bibinfo {volume} {78}},\ \bibinfo {pages} {180410} (\bibinfo
  {year} {2008})}\BibitemShut {NoStop}%
\bibitem [{\citenamefont {Coleman}(2007)}]{coleman2007heavy}%
  \BibitemOpen
  \bibfield  {author} {\bibinfo {author} {\bibfnamefont {P.}~\bibnamefont
  {Coleman}},\ }\href@noop {} {\bibfield  {journal} {\bibinfo  {journal}
  {Handbook of Magnetism and Advanced Magnetic Materials}\ } (\bibinfo {year}
  {2007})}\BibitemShut {NoStop}%
\bibitem [{\citenamefont {Coleman}\ \emph {et~al.}(2005)\citenamefont
  {Coleman}, \citenamefont {Marston},\ and\ \citenamefont
  {Schofield}}]{coleman2005transport}%
  \BibitemOpen
  \bibfield  {author} {\bibinfo {author} {\bibfnamefont {P.}~\bibnamefont
  {Coleman}}, \bibinfo {author} {\bibfnamefont {J.}~\bibnamefont {Marston}}, \
  and\ \bibinfo {author} {\bibfnamefont {A.}~\bibnamefont {Schofield}},\
  }\href@noop {} {\bibfield  {journal} {\bibinfo  {journal} {Physical Review
  B}\ }\textbf {\bibinfo {volume} {72}},\ \bibinfo {pages} {245111} (\bibinfo
  {year} {2005})}\BibitemShut {NoStop}%
\bibitem [{\citenamefont {Wen}(2002)}]{wen2002quantum}%
  \BibitemOpen
  \bibfield  {author} {\bibinfo {author} {\bibfnamefont {X.-G.}\ \bibnamefont
  {Wen}},\ }\href@noop {} {\bibfield  {journal} {\bibinfo  {journal} {Physical
  Review B}\ }\textbf {\bibinfo {volume} {65}},\ \bibinfo {pages} {165113}
  (\bibinfo {year} {2002})}\BibitemShut {NoStop}%
\end{thebibliography}%

\end{document}